\input{aipcheck}

\documentclass[
    ,final            
  ,numberedheadings 
  ]
  {aipproc}

\layoutstyle{6x9}

\begin{document}

\title{X-ray Evolution of SN 1987A}

\author{Bernd Aschenbach}{
  address={Max-Planck-Institut f\"ur extraterrestrische Physik, Giessenbachstra\ss e, 85741 Garching, Germany}
}

%
%

\begin{abstract}
The X-ray observations of SN 1987A over the previous 20 years 
have seen the emergence of soft X-rays from the interaction of 
the explosion shock wave with the ambient medium. This shock wave 
is now interacting strongly with the inner ring and might have passed 
already the highest density regions. The emission can be described 
by thermal models with temperatures of $\sim$0.28 keV and 
$\sim$2.8$\sp{+0.7}\sb{-1.0}$ keV, 
with perhaps some but little change over time. High resolution spectroscopy reveals 
a large variety of shock velocities ranging from a few hundred to many thousand 
km~s$\sp{-1}$. Relative to the elemental abundances prevailing in the LMC 
the inner ring shows an overabundance of Si and S compared to the lighter 
elements and Fe, which suggests that the ring consists of highly 
processed matter dredged up in a  
binary merger event well before the explosion.  
The X-ray lightcurves between 0.5--2 keV and 3--10 keV differ significantly 
in slope, with the latter being much flatter but very similar to the 
radio light curve.  
\end{abstract}

\maketitle


\section{Introduction}

When the star Sanduleak -69$\sp{o}$202 exploded on February 23, 1987, in the Large 
Magellanic Cloud at a distance of just 50 kpc the whole 
astronomical community went into some sort of highly excited state. This 
supernova SN 1987A was considered to be the once-in-a-lifetime event, and 
for the first time the full arsenal of modern days observational tools 
could be used to unravel the secrets of a star's death (section 2), 
and the detailed, 
spatially and temporarily, interaction of shock waves with the ambient 
circumstellar matter (section 3).

\section{Hard X-rays}
 From an X-ray astronomer's point of view the supernova can  
tell us about the explosive creation of matter, in particular radioactive 
material, and its velocity distribution by the appearance of hard 
X-rays. They appear because the $\gamma$-rays from radioactive decay 
($\sp{56}$Ni, $\sp{57}$Ni, $\sp{56}$Co, $\sp{57}$Co, $\sp{44}$Ti, $\sp{22}$Na)  
loose some energy by Compton scattering 
 to become hard X-rays. This goes along with 
 heating the expanding stellar 
envelope. The optical depth was sufficiently low for the first X-rays 
to escape from the debris after about half a year. Both the {\it{Ginga}} 
satellite 
\cite{Dotani87} and the {\it{Roentgen-Observatory}} of the {\it{Kvant-Module}} 
aboard the {\it{Mir-Station}} 
\cite{Sunyaev87,Sunyaev90}
measured the emission, which could be followed 
up to day 835 after the explosion 
 when it began to 
 fade below the sensitivity limit 
of the instruments (c.f. Fig.~1).
The first sighting of the hard X-rays occurred probably on day 143
\cite{Englhauser96}.  
The direct measurement of the 847 keV and 1238 keV $\gamma$-ray lines of 
$\sp{56}$Co and the temporal evolution was successfully performed  
with the spectrometer 
on board of the {\it{Solar Maximum Mission}}
\cite{Matz88}. 
In summary, the theoretical concepts were shown to be all right, and the 
measurements actually put numbers on the amount of radioactive mass. 
Some puzzles still remain, out of which the mixing of radioactive cobalt 
through the supernova's shell with velocities of up to 3000 km s$\sp{-1}$  might 
be mentioned.
 
\begin{figure}
  \includegraphics[bb=150 227 490 670,width=0.75\textwidth,angle=0,clip=]
   {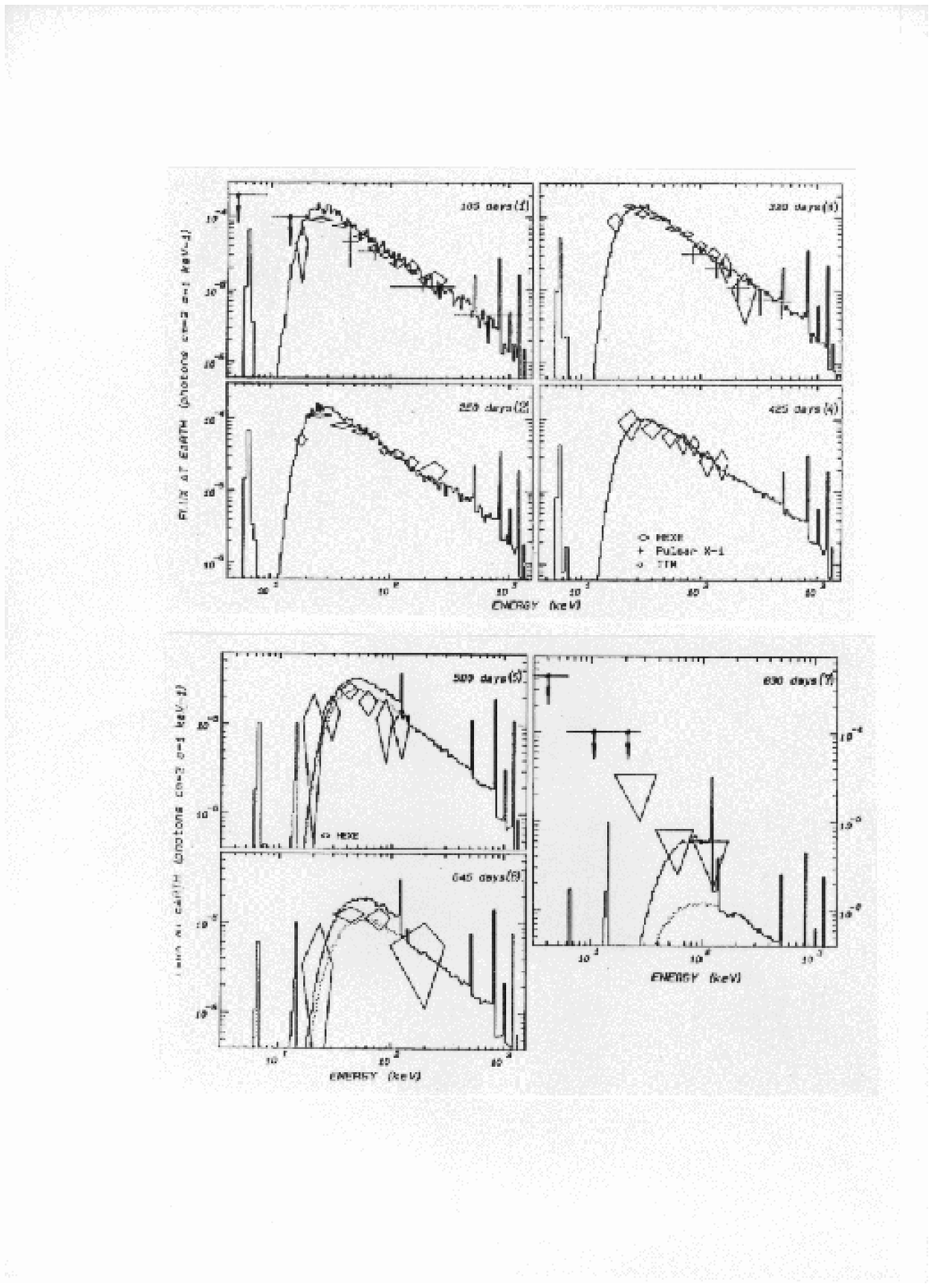}
  \caption{Hard X-ray spectra of SN 1987A from day 185 (upper left) to 
           day 830 (lower right) \cite{Sunyaev90}, comparing measurements and models}
\end{figure}

\section{Soft X-rays}

Soft X-rays with a thermal spectrum were expected to be created by the 
explosion blastwave and the interaction with the ambient medium and eventually by the 
reverse shock with the expanding stellar debris. The first search for 
such emission between 0.5 - 2 keV was carried out on August 24, 1987 with a sounding 
rocket experiment from Australia using an imaging Wolter telescope and 
the prototype of the {\it{ROSAT}} PSPC detector. As Fig.~2 shows, the 
result was fairly disappointing but the upper limit of 1.6$\times$
10$\sp{36}$ erg s$\sp{-1}$ excluded the presence of any dense material around the progenitor 
star including a typical wind of a red supergiant 
\cite{Aschenbach87}. The result was just compatible 
with parameters typical for a blue supergiant wind. Unfortunate to  the 
experimenters was the slightly earlier discovery on archival optical 
plates of the supernova progenitor as blue supergiant.

\begin{figure}
  \includegraphics[bb=90 320 468 680,height=0.38\textheight,angle=0,clip=]
   {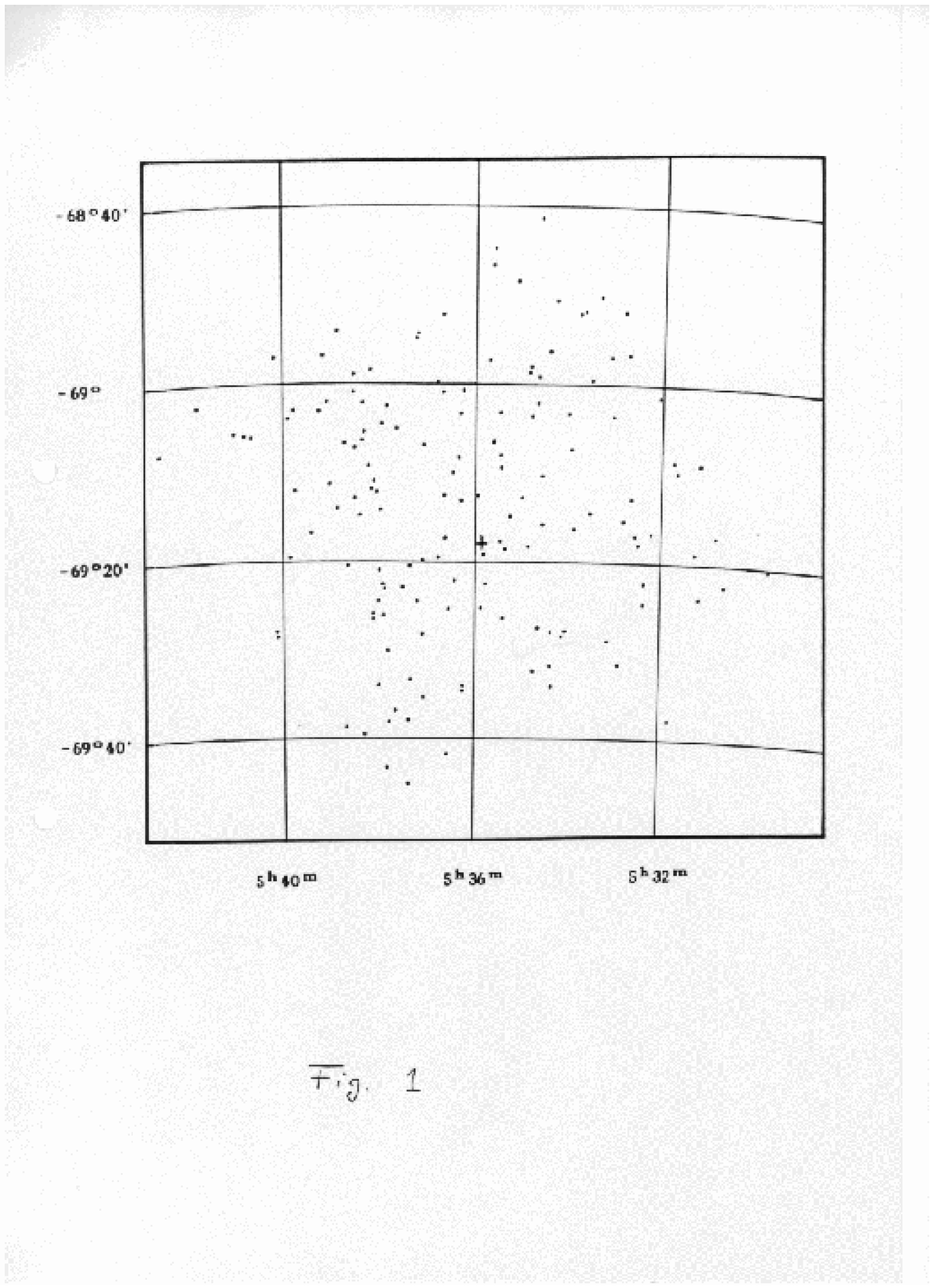}
  \caption{First soft X-ray image of the LMC region centered on SN 1987A 
           (the cross) taken on August 24, 1987
          \cite{Aschenbach87}}
\end{figure}

It took about another 4.5 years until a faint soft X-ray signal from SN 1987A
was detected with {\it{ROSAT}} \cite{Beuermann94,Gorenstein94}. The 
continuous monitoring with {\it{ROSAT}} revealed a slow but steady increase 
in flux over the first few years \cite{Hasinger96} until the end of mission 
of {\it{ROSAT}}.
Measurements were taken up again after the launch of {\it{Chandra}} and later 
with {\it{XMM-Newton}} in 1999 and 2000, respectively. 
The monitoring continued until today 
with fairly regular observations by {\it{Chandra}} 
\cite{Burrows00,Park02,Park04,Park05,Park06}
 and occasionally 
with {\it{XMM-Newton}} \cite{Haberl06}.
 A time  sequence of the beautiful {\it{Chandra}} images can be found in these 
proceedings, see for instance \cite{Park07}.

\par
The analysis of the spectra taken with {\it{ROSAT}}, {\it{Chandra}}, 
{\it{XMM-Newton}} and most recently with {\it{Suzaku}} demonstrate 
that the spectrum is thermal over the entire energy range, 
although the 3--10 keV section of the January 2000 {\it{XMM-Newton}} 
spectrum is also consistent with a power-law. 
The thermal character of the spectra shows that the emission is 
from plasma heated to X-ray temperatures by shock waves. 
The flux as well as the evolution of the lightcurves 
depend on the spatial pre-shock matter density distribution. 
An excellent starting point are the images taken with  
{\it{HST}} and {\it{Chandra}} as well as earlier 
optical observations from the ground. SN 1987A is probably 
embedded in a close to spherical bubble of gas, which might be 
the relic from the winds of the blue and/or red supergiant phase. 
Furthermore, there is the triple ring system with the inner equatorial 
 ring. This ring has 
pre-shock electron 
densities exceeding a few times 10$\sp 4$ cm$\sp{-3}$ \cite{Lundqvist96}, 
which promised to provide a spectacular brightening in soft X-xays when 
the blast wave would hit the ring.
Meanwhile optical and X-ray, as well as radio images of this ring are 
available and the ring 
is actually partially resolved also around its circumference, showing numerous bright 
but unresolved spots in the optical but with less bright regions in between 
\cite{Lawrence00}, which makes it appear as some sort of leaky structure to the blastwave. 
A detailed picture of the environment and the various interactions 
can be found in Michael et al. \cite{Michael03}.

\subsection{Lightcurves}

\begin{figure}[h]
  \includegraphics[
    bb=53 69 462 788,%
   width=0.52\textwidth,angle=-90,clip=]
   {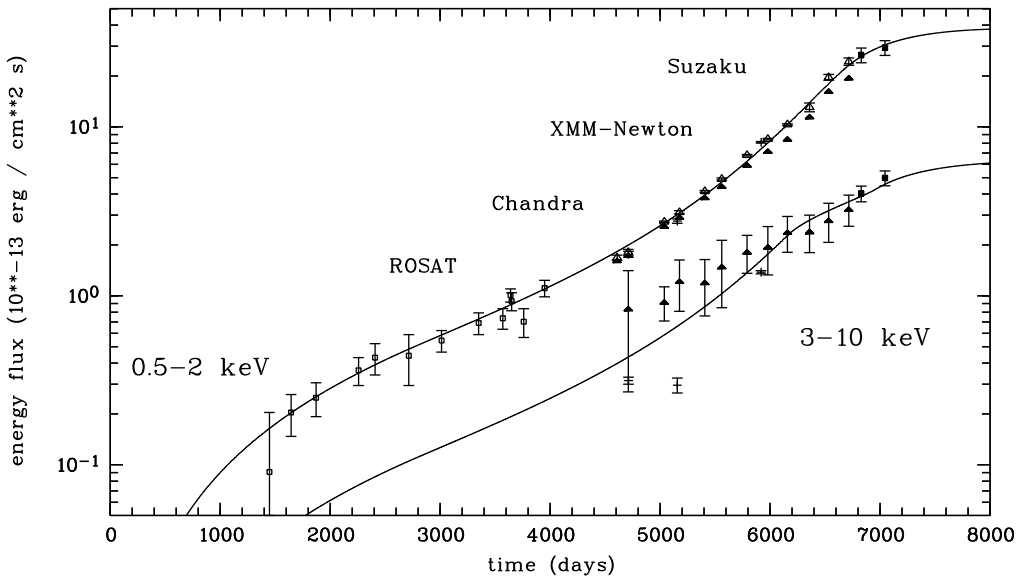}
  \caption{0.5--2 keV and 3--10 keV X-ray lightcurves compiled
    from {\it{ROSAT}}: squares lower left; {\it{Chandra}}: filled
     triangles are original data \cite{Burrows00,Park02,Park04,Park05,Park06},
     open triangles are modified data taken from \cite{Haberl06};
     {\it{XMM-Newton}}: crosses; {\it{Suzaku}}: squares upper right}
\end{figure}

Fig.~3 shows a compilation of all flux data available up to 
the time of this conference for the energy bands of 0.5--2 keV and 3--10 keV, 
respectively. 
The 0.5--2 keV lightcurve with the exception of the two recent {\it{Suzaku}} data 
points has  
been taken from Haberl et al. \cite{Haberl06}. For {\it{Chandra}}
Haberl et al. show the original data published by Park et al. 
\cite{Park02,Park04,Park05,Park06} and a set of slightly corrected data in order to achieve 
a better agreement between {\it{Chandra}} and {\it{XMM-Newton}} in the 
soft band, which I included in Fig.~3. A discussion about the 
appropriateness of this procedure can be 
found in these proceedings \cite{Park07}. 
I might mention that at this point in time there are still uncertainties 
regarding the intercalibration of the instruments including the 
 {\it{Suzaku}}  
data, which are under way to be resolved by the calibration working groups.  
Independent of this slight controversy Fig.~3 shows that the 
shape of the 0.5--2 keV lightcurve differs significantly from that of the 
3--10 keV lightcurve which is less steep. 

\begin{figure}
  \includegraphics[
   width=0.75\textwidth,angle=0,clip=]
   {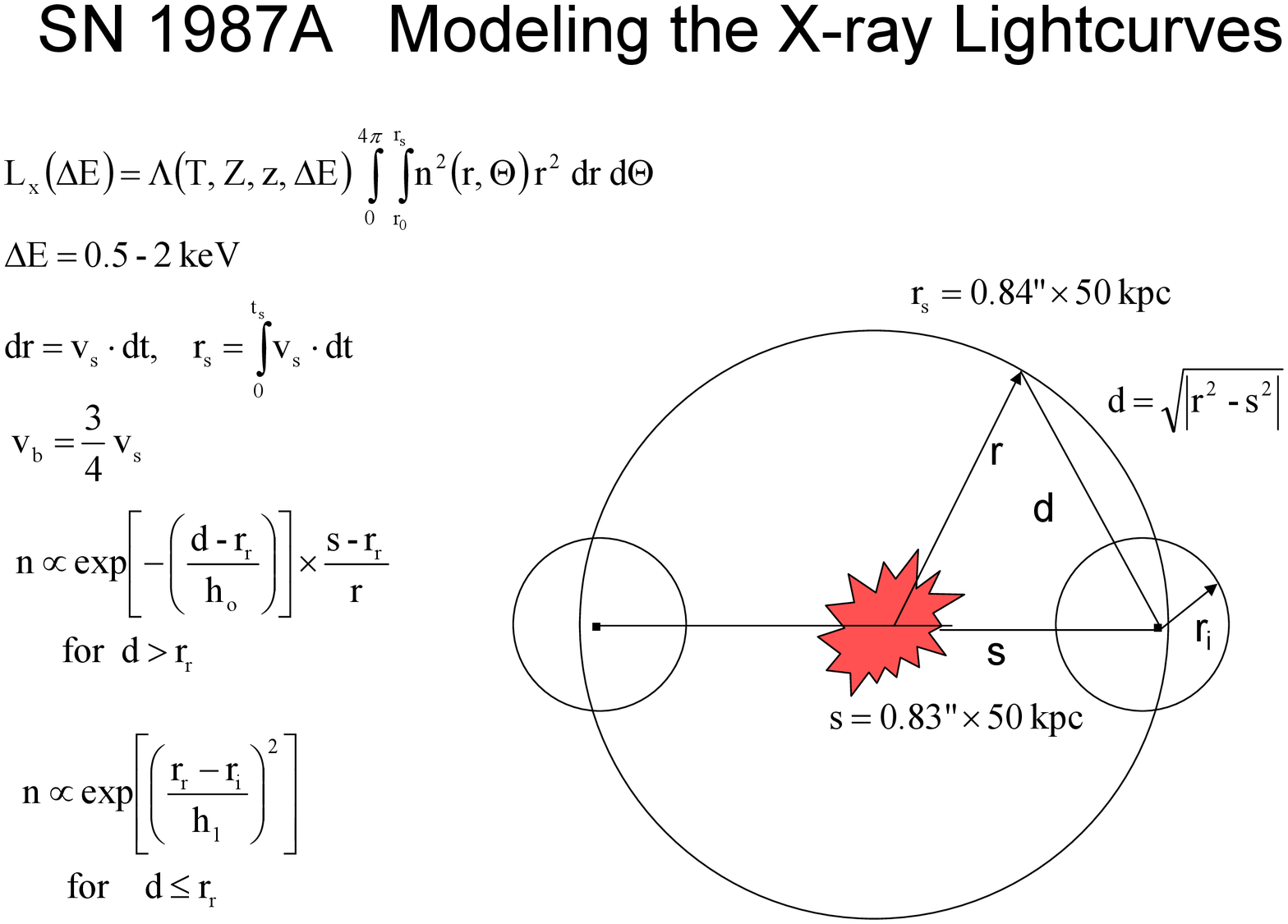}
  \caption{Sketch used to model the matter distribution around 1987A. 
       The picture shows a cross section through the ring. 
       Azimuthal symmetry is assumed}
\end{figure}

The solid lines are the result of a model, which is outlined in Fig.~4. 
The model assumes that the iso-density contours of the ambient medium 
form a circular torus around the explosion site with a circular cross section. 
The density distribution is independent of the azimuth, and perpendicular to 
the iso-density contours it falls off as a Gaussian with increasing distance 
from the torus core. Beyond some distance the Gaussian turns into an 
exponential. The fit to the lightcurve  data provides a maximum pre-shock density 
of 1.2$\times$10$\sp 4$ cm$\sp{-3}$ for the innermost core of the 
ring, which agrees quite nicely with the ring electron density of 
n$\sb{\rm{e}}$~=~(2--4)$\times$10$\sp 4$ cm$\sp{-3}$ derived from the 
early optical/UV measurements \cite{Lundqvist96}. 
For the central 0.1 arcsec diameter ring region the model suggests a mass of 
$\sim$0.065 solar 
masses.  
However, the total mass overrun by the shock wave to date amounts 
to $\sim$0.45 solar masses, 
apparently much more than earlier assumed. 
The model constrains the current position of the shock wave, which has passed the core 
of the ring on day 6900 and is from then on already running in a lower density regime. 
This leads to a flattening of the lightcurve from day 6600 onwards. From now on the 
soft X-ray 
lightcurve is going to climb only very moderately reaching a maximum around the year 
2010 with an increase in flux of less than 50$\%$. A prediction beyond that date is difficult to 
make but a re-brightening 
may happen when the blast wave is going to strike the base of the red supergiant wind, 
if it is not located too far upstream. Another brightening may happen when the stellar ejecta 
are heated by reverse and/or reflected shocks.
 
For the construction of the 3--10 keV lightcurve the same model has been used but those regions  
with densities greater than $\sim$500 cm$\sp{-3}$ have been excluded from the emission, because 
the shock velocity is likely to be too low to produce the required high temperatures. 
\hfill\vfill

\begin{figure}[t]
  \includegraphics[
    bb=53 69 462 788,%
   width=0.50\textwidth,angle=-90,clip=]
   {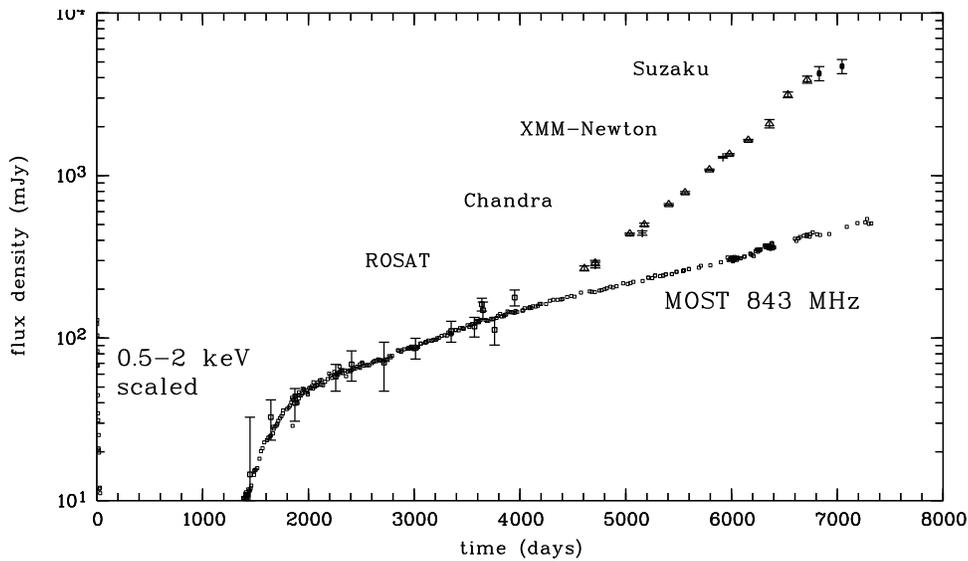}
  \caption{Comparison of the 0.2--2 keV \cite{Haberl06} and the 843 MHz 
      lightcurves \cite{Ball01}, http://www.aip..org/pacs/index.html.
      Note: the X-ray flux density is unlike the radio data not in mJy
      but arbitrarily scaled to match best the radio data. 
      Furthermore, the most 
recent radio observations of 2007 shown may be more unreliable than 
indicated by the formal errors.}
\end{figure}

\begin{figure}
  \includegraphics[
    bb=53 69 462 788,%
   width=0.50\textwidth,angle=-90,clip=]
   {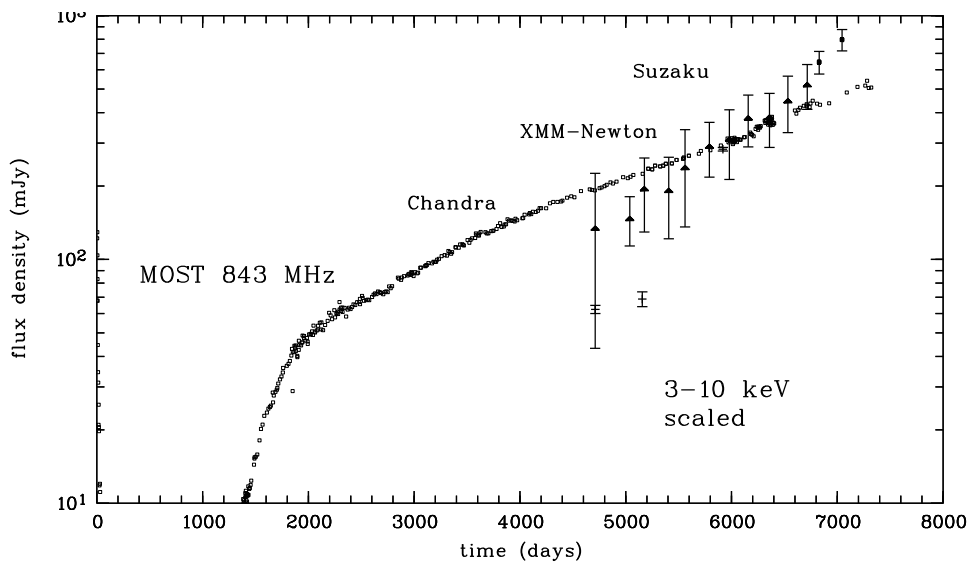}
  \caption{Comparison of the 3--10 keV \cite{Haberl06} and the 843 MHz
      lightcurves \cite{Ball01}, http://www.aip..org/pacs/index.html. 
      Note: the X-ray flux density is unlike the radio data not in mJy
      but arbitrarily scaled to match best the radio data.
      Furthermore, the most
recent radio observations of 2007 shown may be more unreliable than
indicated by the formal errors.}
\end{figure}

\par

Fig.~5 and Fig.~6 repeat the X-ray lightcurves for the
 0.5--2 keV and 3--10 keV bands, respectively,
and each of which is compared with the most recent MOST radio lightcurve at 843 MHz
(see http://www.physics.usyd.edu.au/ioa/Main/SN1987A) \cite{Ball01}.
Up to day $\sim$4000 the radio data agree with the soft X-ray lightcurve surprisingly well, but they don't
follow the rapid increase 
of the X-rays later on; instead the radio data continue to rise but at a much slower but almost
constant rate (Fig.~5). The agreement with the hard X-ray lightcurve (Fig.~6) is much better, although the
{\it{Chandra}} data supplemented by the more recent {\it{Suzaku}} data indicate a faster, although constant rise as
well. But, as  said before, we need to look into the X-ray calibration issues.
 Nevertheless, it seems unavoidable to conclude that both hard X-rays  as well as  
843 MHz emission are lacking in the highest density regions of the ring. 
For the X-rays it means that the shock velocity is too low and for the radio emission it means that 
the acceleration of electrons to GeV energies and/or the compression of magnetic fields is less efficient 
or even suppressed.
Images with significantly improved angular resolution, which resolve the optical bright (hot) spots 
would be useful to follow this issue.   

\subsection{Spectra}

X-ray spectra have been recorded since 1999 by {\it{Chandra}} and 
{\it{XMM-Newton}} both at medium spectral resolution with the X-ray 
CCD-detectors 
\cite{Burrows00,Park02,Park04,Park05,Park06,Haberl06} 
and high resolution with the transmission grating {\it{LETG}} 
on {\it{Chandra}} \cite{Zhekov05,Zhekov06} and 
the reflection grating spectrometers {\it{RGS1 \& RGS2}} on 
{\it{XMM-Newton}} \cite{Haberl06}. 
Recently, the CCD-detectors aboard of {\it{Suzaku}} contributed further 
spectra. 
For a view of the {\it{Chandra}} spectra I refer to the paper of Park et al. 
\cite{Park07} in these proceedings. 
Fig.~7 shows the {\it{XMM-Newton EPIC-pn}} spectra taken in January 2000, April 2001 and 
May 2003. The {\it{RGS1 \& RGS2}} grating spectra of the May 2003 observation are shown in Fig.~8.  

For the analysis of the spectra different models for the purely 
thermal emission have been used. The {\it{Chandra}} data have been  
compared with predictions of a two-component plane 
shock model whereas {\it{XMM-Newton}} and {\it{Suzaku}}
data were subjected to an analysis with a two-temperature model with 
collisional ionization equilibrium and non-ionization equilibrium components.  
Despite the different approaches the results are pretty close to each 
other as far as the low temperature of 0.28 keV on average is concerned. In contrast 
differences in the value 
of the high temperature are fairly large when comparing {\it{Chandra}}, 
{\it{XMM-Newton}} and {\it{Suzaku}}. One may quote a temperature of 2.8$\sp{+0.7}\sb{-1.0}$ keV; a
clear temporal trend is not apparent, although some excursions down and up again after 
day $\sim$ 6000 may not be excluded.  
The low temperature started off at $\sim$0.23 keV in 1999 and has now 
reached at value of $\sim$0.33 keV, which would correspond to an increase 
of $\sim$100 eV~(7 years)$\sp{-1}$. This increase may indicate that the heating of the 
electrons had become more efficient when the shock wave has entered the high 
density regions of the ring. 

\begin{figure}
  \includegraphics[
   width=0.54\textwidth,angle=-90,clip=]
   {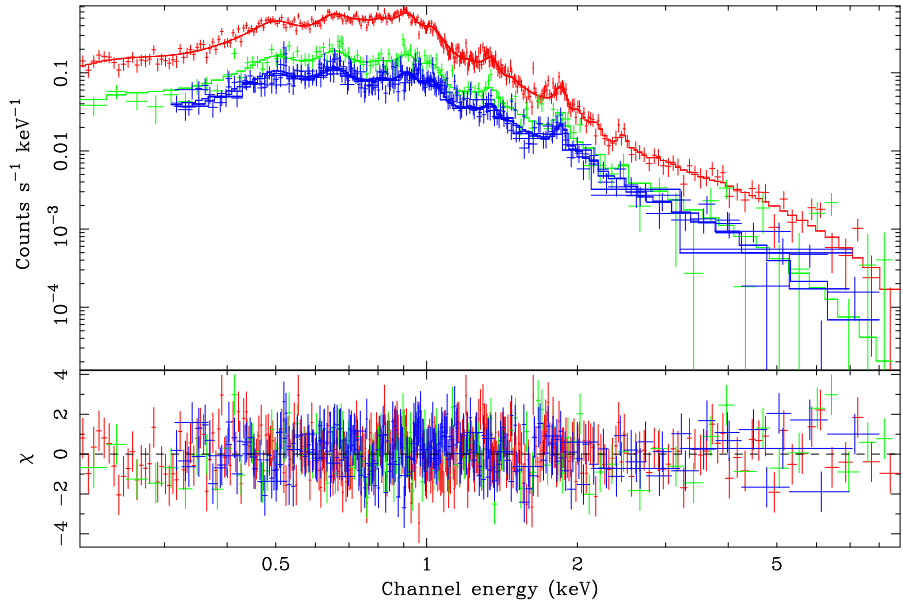}
  \caption{{\it{XMM-Newton EPIC-pn}} spectra of SN 1987A from
      January 2000, April 2001 and May 2003 (top panel) and
      the best fit residuals (bottom panel) \cite{Haberl06}}
\end{figure}

\begin{figure}
  \includegraphics[
   width=0.54\textwidth,angle=-90,clip=]
   {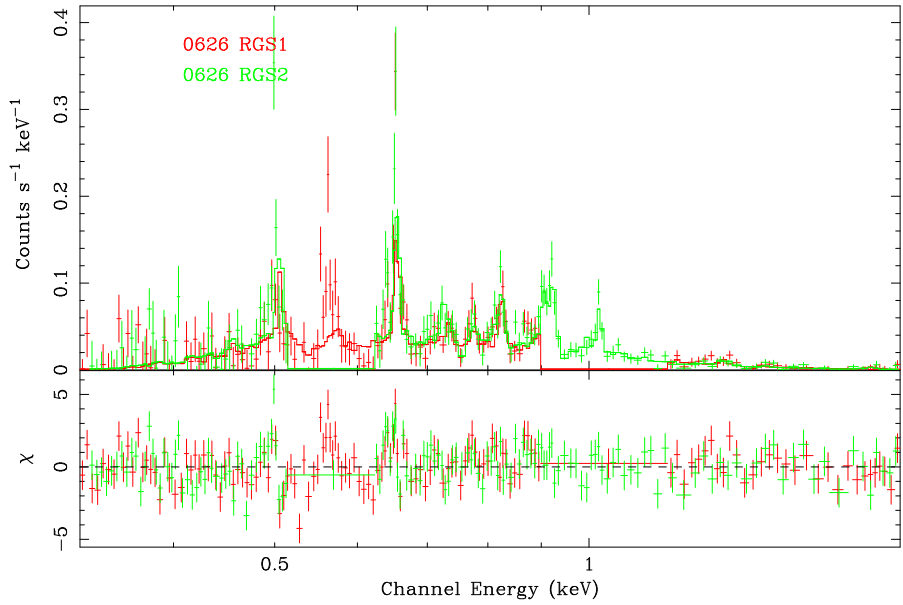}
  \caption{{\it{XMM-Newton RGS}} spectra of SN 1987A from
      May 2003 \cite{Haberl06}. The most prominent peaks are
      associated with N~VII, O~VII, O~VIII, Fe~XVII, O~VII, Fe~XVII,
      Ne~IX and Ne~X with increasing energy, i.e. from left to right}
\end{figure}

The high resolution data of the gratings could be used to assess the presence of numerous 
high ionization emission lines between 0.5--2 keV, such as N~VII, O~VII, O~VIII, Fe~XVII, 
Ne~IX, Ne~X, Mg~XI, Mg~XII, Si~XIV,~XV,~XVI (Fig.~8). Their shapes and shifts have revealed a substantial 
range of shock velocities running from a few hundred to many thousand km~s$\sp{-1}$, which basically 
reflects the complex density structure of the ambient matter. 
I note that the blastwave speed averaged over the past 20 years is 10500 km~s$\sp{-1}$, though.

\subsection{Elemental abundances}

Both the medium resolution and the high resolution data have been used 
to compile elemental abundances, the distribution of which is shown in 
Fig.~9. The abundance of each element covered by the data and measured 
relative to the corresponding solar values have been computed by averaging the 
corresponding data from each contributing instrument in an unweighted 
way. Then, this averaged abundance has been normalized for each element 
to a so-called 
typical abundance of the LMC given by Hughes et al. \cite{Hughes98}. 
These have been derived from {\it{ASCA}} observations of 
middle-aged supernova remnants in the LMC.
Fig.~9 shows that the elements of N, O, Ne, Mg and Fe in the 
inner ring and in the region enclosed currently by the blastwave 
have about the 
same abundance as are present in the LMC. But Si, S and 
possibly Ni (uncertain, because this is a {\it{Suzaku}} measurement 
not yet confirmed) turn out to be much higher in abundance. There 
appears to be an enrichment of these elements in the inner ring by about a 
factor of three. 

\begin{figure}
  \includegraphics[
    bb=45 80 558 614,
   width=0.60\textwidth,angle=-90,clip=]
   {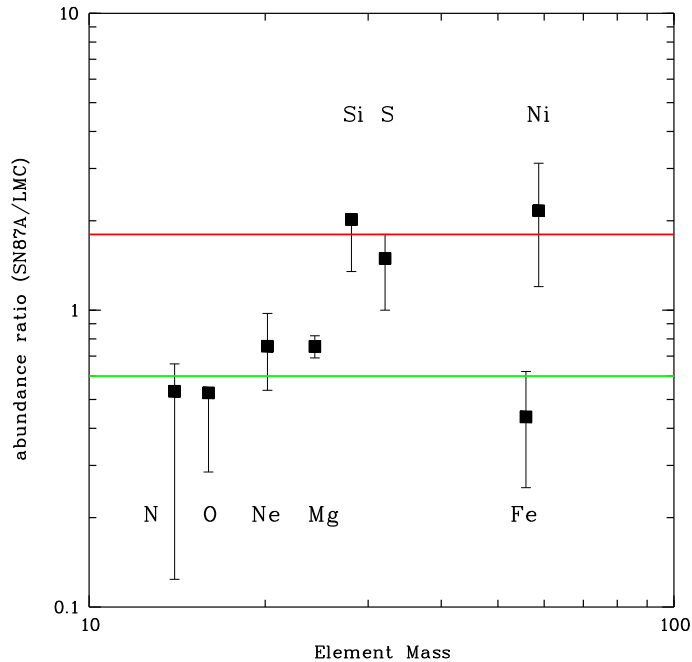}
  \caption{Ratio of elemental abundances in the inner ring and 
      the average LMC. Silicon and sulphur, for instance, 
      are overabundant in the inner ring in relation to the lighter 
      elements and Fe. The ratio is defined as 
      (X/H)$\sb{\rm{ring}}$/(X/H)$\sb{\rm{LMC}}$; X stands for some element 
     and H for hydrogen. The absolute ratio shown depends on the abundance of H 
      in the inner ring and that prevailing in the LMC and may scale 
      with this H-ratio. The relative abundance ratios do not}
\end{figure}

In terms of absolute abundances the numbers have to 
be taken with some care because they depend on how much hydrogen is 
involved in either the inner ring or the LMC in general, but the 
ratio of the average of the two groups remains to be the same. 
It looks as if the inner ring contains preferentially highly processed 
material. It is unlikely that this has been expelled from the outermost 
layers of the progenitor star or that the ring  formed from  
interaction of winds. In contrast, it seems to favour 
a binary merger as the event which created the inner ring or even 
the entire ring system as suggested earlier by Morris \& Podsiadlowski
\cite{Morris06,Morris07}.

\begin{theacknowledgments}
I like to thank G\"unther Hasinger and Masayuki Itoh for providing the {\it{Suzaku}} X-ray data prior to 
publication. I am grateful to Richard Hunstead for giving permission of using the MOST radio 
data: The MOST is operated by the University of Sydney and supported in part by grants 
from the Australian Research Council.
\end{theacknowledgments}

\end{document}